\documentclass{elsart}
\usepackage{epsfig}

\def\xs{cross-section} \def\xss{cross-sections}
\def\beq{\begin{equation}} \def\eeq{\end{equation}}
\def\beqa{\begin{eqnarray}} \def\eeqa{\end{eqnarray}}
\def\bce{\begin{center}}  \def\ece{\end{center}}
\def\bfig{\begin{figure}}  \def\efig{\end{figure}}
\def\bit{\begin{itemize}}    \def\eit{\end{itemize}}
\def\ben{\begin{enumerate}}    \def\een{\end{enumerate}}
\def\npb{Nucl. Phys. {\bf B}} \def\plb{Phys. Lett. {\bf B}}
\def\prd{Phys. Rev. {\bf D}}
\def\prp{Phys. Rep.} \def\prl{Phys. Rev. Lett.}

\begin{document}

\newlength{\caheight}
\setlength{\caheight}{12pt}
\multiply\caheight by 7
\newlength{\secondpar}
\setlength{\secondpar}{\hsize}
\divide\secondpar by 3
\newlength{\firstpar}
\setlength{\firstpar}{\secondpar}
\multiply\firstpar by 2

\hfill
\parbox[0pt][\caheight][t]{\secondpar}{
  \rightline
  {\tt \shortstack[l]{
    FNT/T-99/08\\
    CERN-TH/99-125
  }}
}

\begin{frontmatter}

\title{Top-quark physics in six-quark final states at the Next
       Linear Collider}

\author{F.~Gangemi$^{1,2}$, G.~Montagna$^{1,2}$, M.~Moretti$^{3,4}$,}
\author{O.~Nicrosini$^{2,1}$ and F.~Piccinini$^{2,1}$}

\address{$^1$Dipartimento di Fisica Nucleare e Teorica, Universit\`a di Pavia,
         via A.~Bassi 6, Pavia, Italy}
\address{$^2$Istituto Nazionale di Fisica Nucleare, Sezione di Pavia, via
 A.~Bassi 6, Pavia, Italy}
\address{$^3$CERN - Theory Division, Geneva, Switzerland}
\address{$^4$Dipartimento di Fisica, Universit\`a di Ferrara and INFN, Sezione
 di Ferrara, Ferrara, Italy}

\begin{abstract}

The processes of six-quark production with one $b\bar b$ pair are studied
by means of a complete tree-level electroweak calculation. The  top-quark
signal is examined: the importance of electroweak backgrounds, of the order of
10$\%$ above the $t\bar t$ threshold and of about 30$\%$ of the purely
electroweak signal at threshold, is
further stressed by studying the dependence of the \xs\ at threshold
on the Higgs mass in the range between 100 GeV and 185 GeV, and finding
variations of the order of 10$\%$. In the study of some event-shape variables,
a strong effect of initial-state radiation is found, in particular for the
thrust distribution, which is studied for several centre-of-mass energies at
the TeV scale. The effectiveness of cuts on the thrust for isolating QCD
backgrounds, as pointed out by some authors, is confirmed also in the presence
of electroweak backgrounds and initial-state radiation.

\end{abstract}


\end{frontmatter}

\section{Introduction}
\label{sect:intro}

Many important signals to be studied at future high energy colliders,
NLC~\cite{lc} and LHC~\cite{lhc}, will have a large number of particles in
the final state. In particular, the processes with six
fermions in the final state will be of great importance for several tests of
the Standard Model, such as the studies of top-quark and electroweak
gauge bosons, as well as the search for an intermediate-mass Higgs boson.
Such processes are already of great interest at the Tevatron
collider~\cite{Fermilab} in connection with top-quark physics.
Theoretical studies of six-fermion ($6f$) processes by means of complete
tree-level calculations have only very recently appeared in the
literature~\cite{kek,to1,sixfzpc,gmmnp97,6fhiggs,keystone}, where top-quark
 physics, Higgs physics and $WWZ$ production have been addressed.

All these studies clearly demonstrate that complete calculations are important
for a precise determination of the cross-sections, and for the development
of reliable event generators, whenever accurate evaluations of interference,
off-shellness and background effects,
 as well as spin correlations, are important.

In view of the precision measurements of the top-quark properties we have
to analyse, among the $6f$ signatures, the ones containing a $b\bar b$ pair
and two charged currents, as the top-quark decays almost exclusively into
a $W$ boson and a $b$ quark.
Semi leptonic signatures have already been considered in
refs.~\cite{kek,to1,4jln}. It is then of great interest to carefully
evaluate the size of the totally hadronic, six-quark ($6q$) contributions to
integrated cross-sections and distributions as well as to determine
their phenomenological features.
The aim of the present study is to make a first quantitative analysis in this
context, for what concerns the full set of electroweak contributions to a class
of $e^+e^-$ annihilation processes related to top-quark physics. Special
emphasis will be given to the determination and to the analysis of the topology
of the events considered, so as to characterize them, as far as possible, 
against
the QCD backgrounds.

Looking at an experimental situation where the $b$-tagging technique can be
applied, it is meaningful to distinguish the
$6q$ final states containing one $b\bar b$ pair, of the form
$b\bar b q\bar q' q''\bar q'''$, from those containing two or three
$b\bar b$ pairs, respectively of the form $b\bar b b\bar b q\bar q$ and
$b\bar b b\bar b b\bar b$. The last two kinds of processes are not relevant
to top-quark production, as they contain no charged currents.
\begin{table}[h]
\bce
\begin{tabular}[b]{||p{2.6truecm}|p{2.6truecm}|p{2.6truecm}||}
\hline
%\cline{1-3}
CC only & CC and NC & NC only\\
\hline
$b\bar bu\bar d\bar cs$ & $b\bar bu\bar d\bar ud$ & $b\bar bu\bar us\bar s$,
$b\bar bc\bar cd\bar d$ \\

$b\bar b\bar udc\bar s$ & $b\bar bc\bar s\bar cs$ & $b\bar bu\bar uu\bar u$,
$b\bar bc\bar cc\bar c$ \\

                        &                         & $b\bar bd\bar dd\bar d$,
$b\bar bs\bar ss\bar s$ \\

                        &                         & $b\bar bu\bar uc\bar c$ \\
                        &                         & $b\bar bd\bar ds\bar s$ \\
\hline
\end{tabular}
\caption{\small Six-quark final states with one $b\bar b$ pair. The
notations CC (charged currents) and NC (neutral currents) refer to the currents
formed by the quark flavours other than $b$.}
\label{tab:1bb}
\ece
\end{table}

 In the present study the signatures with one $b\bar b$ pair are
considered and the full set of purely electroweak contributions is taken into
account. These processes can be further divided into three subsets (although in
realistic predictions they cannot be treated separately), which are shown
in Table~\ref{tab:1bb}: concerning the quark flavours other than $b$,
only charged currents are involved in the first subset, both charged
and neutral currents in the second one, and only neutral currents in the
third one.
The total number of tree-level Feynman diagrams involved in the complete
electroweak calculation amounts to several hundreds.
Such a complexity is unavoidable, as  will be shown, if an accuracy of $1\%$
is to be reached.
 The diagrams with top-quark production, which will be referred to as
signal diagrams, are shown in Fig.~\ref{fig:topfd}. They contribute to the
processes in the first two columns of Table~\ref{tab:1bb}, but not to those in
the third.
\bce
\bfig
\bce
\epsfig{file=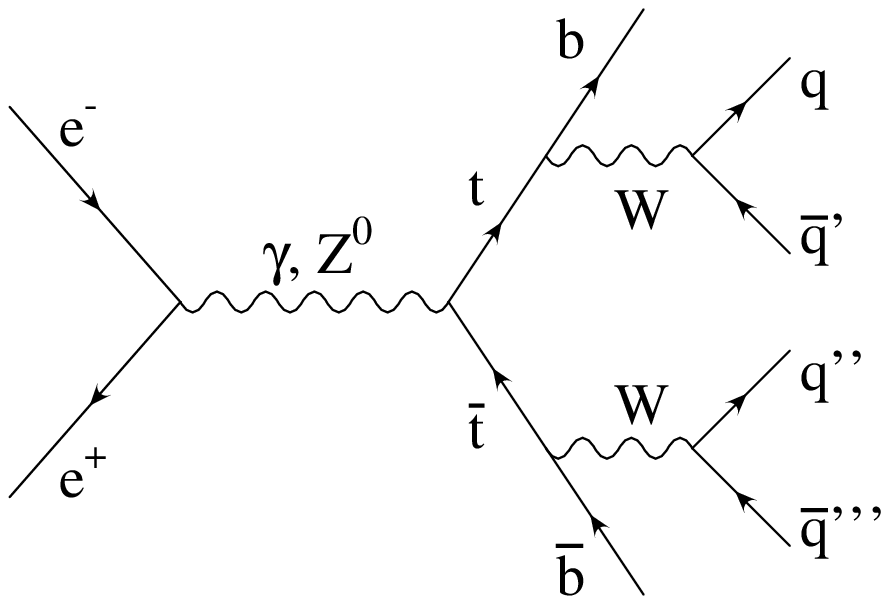,height=4.6cm,width=6.9cm}
\caption{\small The two Feynman diagrams with $t\bar t$ production.}
\label{fig:topfd}
\ece
\efig
\ece
 All the processes receive contributions from diagrams of Higgs production,
of which the leading ones are illustrated in Fig.~\ref{fig:higgsfd}. The
relevance of such contributions depends on the Higgs mass and on the 
centre-of-mass (c.m.) energy: the dominant decay mode is $H\to b\bar b$ for low Higgs
masses ($m_H\le 130$-$140$ GeV) and $H\to VV$ ($V=W,Z$) for high Higgs masses.
The predictions can thus be expected to depend on the Higgs mass.
\bce
\bfig
\bce
\epsfig{file=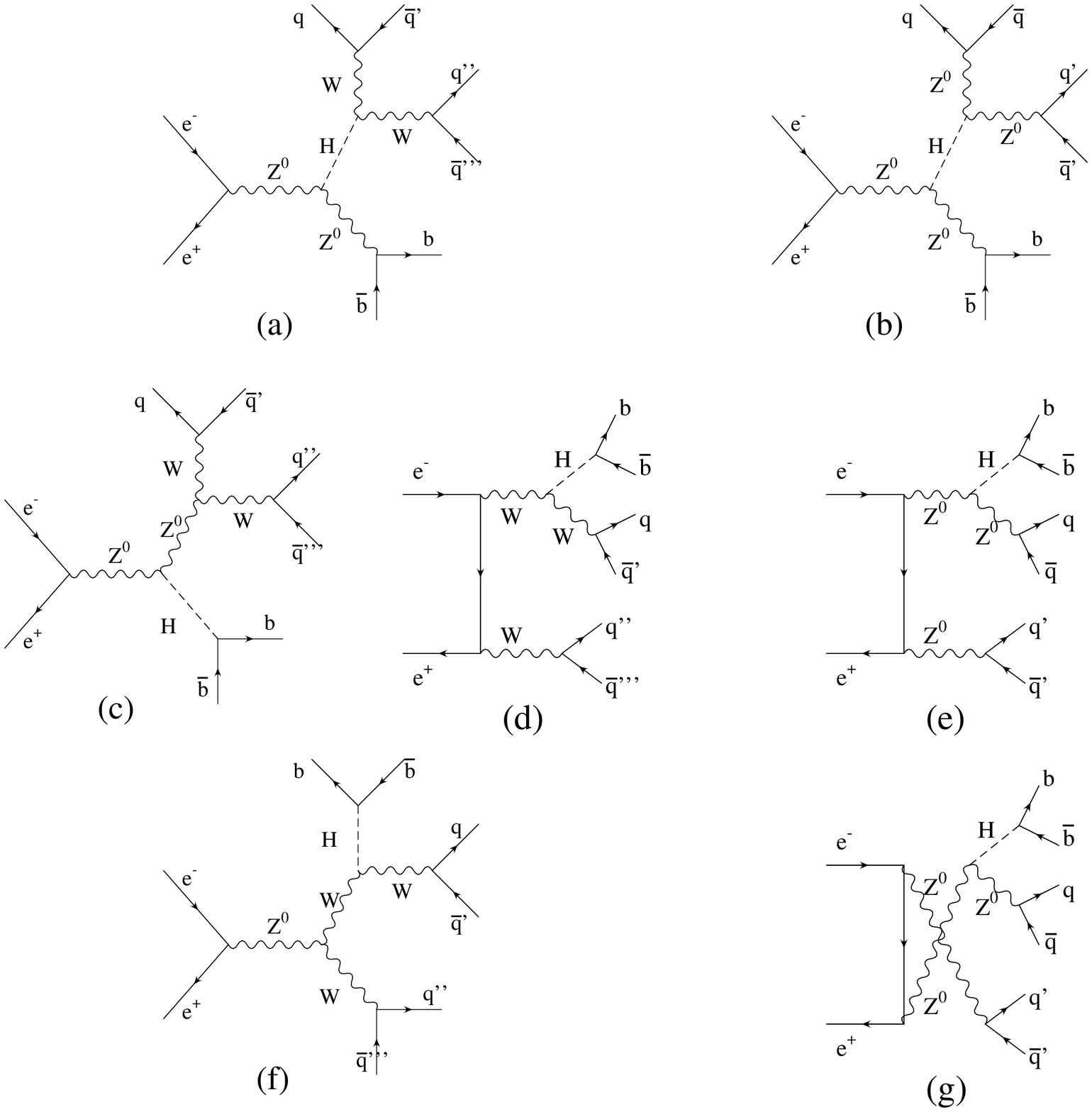,height=13.7cm,width=13.7cm}
\caption{\small Leading Higgs contributions, divided into charged-current
 terms, on the left, and neutral-current terms, on the right (where charged and
 neutral currents are referred to quark flavours other than $b$).
 Other diagrams, obtained by attaching the Higgs line to the other possible
 gauge-boson line in the diagrams $(d)$, $(e)$, $(f)$, and $(g)$, are
 understood.}
\label{fig:higgsfd}
\ece
\efig
\ece
As is well known, the behaviour of the cross-section near the threshold for
$t\bar t$ production is characterized by strong interaction effects that give
a sizeable modification with respect to the purely electroweak prediction.
Such effects are treated in the literature~\cite{threshold,teub}, and are not
included in the calculations presented in this paper.
~\footnote{Theoretical calculations of radiative corrections
to $t\bar t$ production are also present in the literature, as recently reviewed
in ref.~\cite{teub}.} Results at
energies around the threshold are shown, so as to give a thorough analysis
of the electroweak contribution.
Some of the QCD backgrounds to the signatures considered in the present study
have been evaluated in ref.~\cite{sixj}, and their topology has been studied
by means of event-shape variables.
One of the objectives of the present work is to characterize the topology
of the complete electroweak contributions in order to help  finding
appropriate selection criteria to reduce as far as possible the QCD backgrounds
studied in ref.~\cite{sixj}.
The analysis performed in the present work, together with the other studies
 in the literature so far, should give a complete picture of electroweak
contributions to $6f$ processes relevant to top-quark physics at NLC.

The paper is organized as follows: in Section~\ref{sect:calc} the computing
procedure is briefly described; in Section~\ref{sect:results} the numerical
results, including integrated cross-sections and various distributions, are
presented and discussed; Section~\ref{sect:concl} is devoted to our
conclusions.

\section{Calculation}
\label{sect:calc}

The numerical results have been obtained by means of
a procedure analogous to the one adopted in ref.~\cite{6fhiggs}, where the
interested reader can find some technical details that will be omitted here.
The computer program already used in ref.~\cite{6fhiggs}, which is based on
ALPHA~\cite{alpha} for the matrix element calculation and on an evolution
of HIGGSPV/WWGENPV~\cite{higgspv,wwgenpv} for the Monte Carlo integration
and event generation, has been adapted, in the multichannel importance sampling,
to include some new diagram topologies, such as those in Figs.~\ref{fig:topfd}
and \ref{fig:higgsfd}.

The \xs\ is calculated according to the formula
\beq
\label{eq:2.1}
  \sigma =\int dz_1dz_2D_{BS}(z_1,z_2;s)\int dx_1dx_2D(x_1,s)D(x_2,s)
  d[PS]\frac{d\hat\sigma }{d[PS]}\ ,
% (z_1,z_2;x_1,x_2;s)
\eeq
where initial-state radiation ({\em ISR})~\cite{sf} and beamstrahlung
({\em BS})~\cite{circe} are included by means of the structure functions $D(x,s)$ and
$D_{BS}(x,s)$, respectively; $d\hat\sigma/d[PS]$ is the differential
\xs\ at the partonic level, and $d[PS]$ is the six-body phase-space measure.
The program may be used to generate unweighted events as well.

The input parameters are $G_\mu$, $M_W$, $M_Z$, the top-quark mass
$m_t=175$ GeV, and the $b$-quark mass $m_b=4.3$ GeV; all the other fermions
are treated as massless. The widths of the $W$ and $Z^0$ bosons and of the
top-quark and all the couplings are calculated at tree level. The 
Higgs-boson width includes
the $h\to \mu\mu,\tau\tau,cc,bb$, the $h\to gg$~\cite{hwg} and the
two-vector-boson~\cite{kniel} channels. The CKM matrix used is 
exactly diagonal.
The propagators of unstable particles have
denominators of the form $p^2 - M^2 + i\Gamma M$ with fixed widths.
The validity of this choice for minimizing possible gauge violations has been
discussed in ref.~\cite{6fhiggs}, where the final states
$q\bar ql^+l^-\nu\bar\nu$ were considered. In that paper, for the $SU(2)$
invariance, the fudge-factor method has been used to check the numerical
results: apart from the well-known problems of the fudge scheme,
{\em i.e.} the mistreatment of non-resonant diagrams close to the resonances,
no deviation has been found in the total \xs\, up to the numerical
accuracy considered.
In order to check $U(1)$ invariance, the matrix element has been calculated
with different forms of the photon propagator obtained by varying the gauge
parameter;  the results were found to be stable up to numerical
precision. The same analysis carries on to the present study and
gauge-violation effects are estimated to be numerically negligible.

The colour algebra, not implemented in the version of ALPHA that has been
employed here, has been performed by summing the different processes with proper
weights. As an example of this, the process $e^+e^-\to b\bar bu\bar d\bar ud$
may be considered: the colour amplitude, in the case of purely electroweak
contributions, can be written in the form
\beq
\label{eq:2.2}
A = \left(a_1\delta_{i_1i_2}\delta_{i_3i_4}+
    a_2\delta_{i_1i_3}\delta_{i_2i_4}\right)\delta_{jk}\ ,
\eeq
where the colour indices $i_1,i_2,i_3,i_4,j$ and $k$ refer to the $u,\bar d,
\bar u,d,b$ and $\bar b$ quarks respectively. The squared modulus summed over
colours is then
\beq
\label{eq:2.3}
\sum_{col}|A|^2 = N_c^3|a_1|^2 + N_c^2(a_1a_2^*+a_1^*a_2) + N_c^3|a_2|^2\ .
\eeq
The amplitude given by ALPHA is instead
\beq
\label{eq:2.4}
{\mathcal A} = a_1+a_2\ .
\eeq
Thus one cannot use an overall factor to obtain eq.~(\ref{eq:2.3}) from
eq.~(\ref{eq:2.4}). In order to disentangle the various terms in
eq.~(\ref{eq:2.3}), it is useful to notice that, with the quark masses adopted
here and with a diagonal CKM matrix, the first term in the right-hand side of
eq.~(\ref{eq:2.2}) is equal to the amplitude $A'$ of the process
$e^+e^-\to b\bar bu\bar d\bar cs$, and the second term is equal to the
amplitude $A''$ of the process $e^+e^-\to b\bar bu\bar s\bar us$. Similarly, the
two colourless amplitudes $\mathcal{A}'$ and $\mathcal{A}''$ of these processes
are equal to the first and to the second term, respectively, in the right-hand
side of eq.~(\ref{eq:2.4}). Thus the following relation is valid:
\beqa
\label{eq:2.5}
\sum_{col}\left(|A|^2 + |A'|^2 + |A''|^2\right)=
N_c^2\left(|{\mathcal A}|^2 + (2N_c-1)(|{\mathcal A}'|^2+|{\mathcal A}''|^2)
\right)\ .
\eeqa
Other situations are treated in a similar way, and the correct colour weights
are thus obtained in the sum over the whole class of processes considered.

\section{Numerical results and discussion}
\label{sect:results}

In this section the numerical results, including both integrated \xss\ and
distributions, are shown.
In all the calculations the invariant masses of the $b\bar b$ pair and of all
the pairs of quarks other than $b$ and $\bar b$ are required to be greater
than 10 GeV.
 The results presented below are obtained, unless otherwise stated, by summing
over all the processes listed in Table~\ref{tab:1bb}.

\subsection{Integrated cross-sections}
\label{subsect:xs}
As a first step, the total cross-section, resulting from all the tree-level
diagrams for the processes in Table~\ref{tab:1bb}, has been calculated in the
Born approximation at energies from 340 to 800 GeV. Two values of Higgs mass
have been considered, $m_H=100, 185$ GeV, so as to study the dependence of the
results on $m_H$ in the intermediate range. The numerical errors are always
below $1\%$ and in particular above the $t\bar t$ threshold they are kept at
$0.2-0.3\%$ level.

In Fig.~\ref{fig:f185sig} the full cross-section for $m_H = 185$ GeV is
compared with the signal, defined as the contribution of the two diagrams of
$t\bar t$ production of Fig.~\ref{fig:topfd}, summed over the four processes
to which they contribute (see the first two columns of Table~\ref{tab:1bb});
the signal is shown both in the Born approximation and with {\em ISR} switched on.

The difference between the full and the signal curve is dominated by
Higgs-strahlung contributions (diagrams $(a)$ and $(b)$ of
Fig.~\ref{fig:higgsfd}) at low energy, while other backgrounds are
important at high energy, coming from all the processes in Table~\ref{tab:1bb}
and, for a little amount, from the interference of the signal diagrams with the
other contributions in the charged-current and mixed processes.
The electroweak background effects, which are in the range $5-10\%$ above the
threshold, amount to around 30$\%$ at threshold; they 
  are much greater below the
threshold, where the signal is suppressed with respect to the background (the
ratio background/signal is of 2.5 at 340 GeV).

The radiative effects strongly suppress the \xs\ in the low-energy region,
where it grows rapidly, as they reduce the effective c.m. energy; with
increasing energy, the curve with {\em ISR} comes to cross the one in the Born
approximation, as a consequence of the onset of the opposite behaviour of the
Born term, which, above the threshold, starts decreasing. It can be observed
that at 500 GeV the enhancement due to the background is of the same order as
the lowering given by the {\em ISR}.
\bce
\bfig
\bce
\epsfig{file=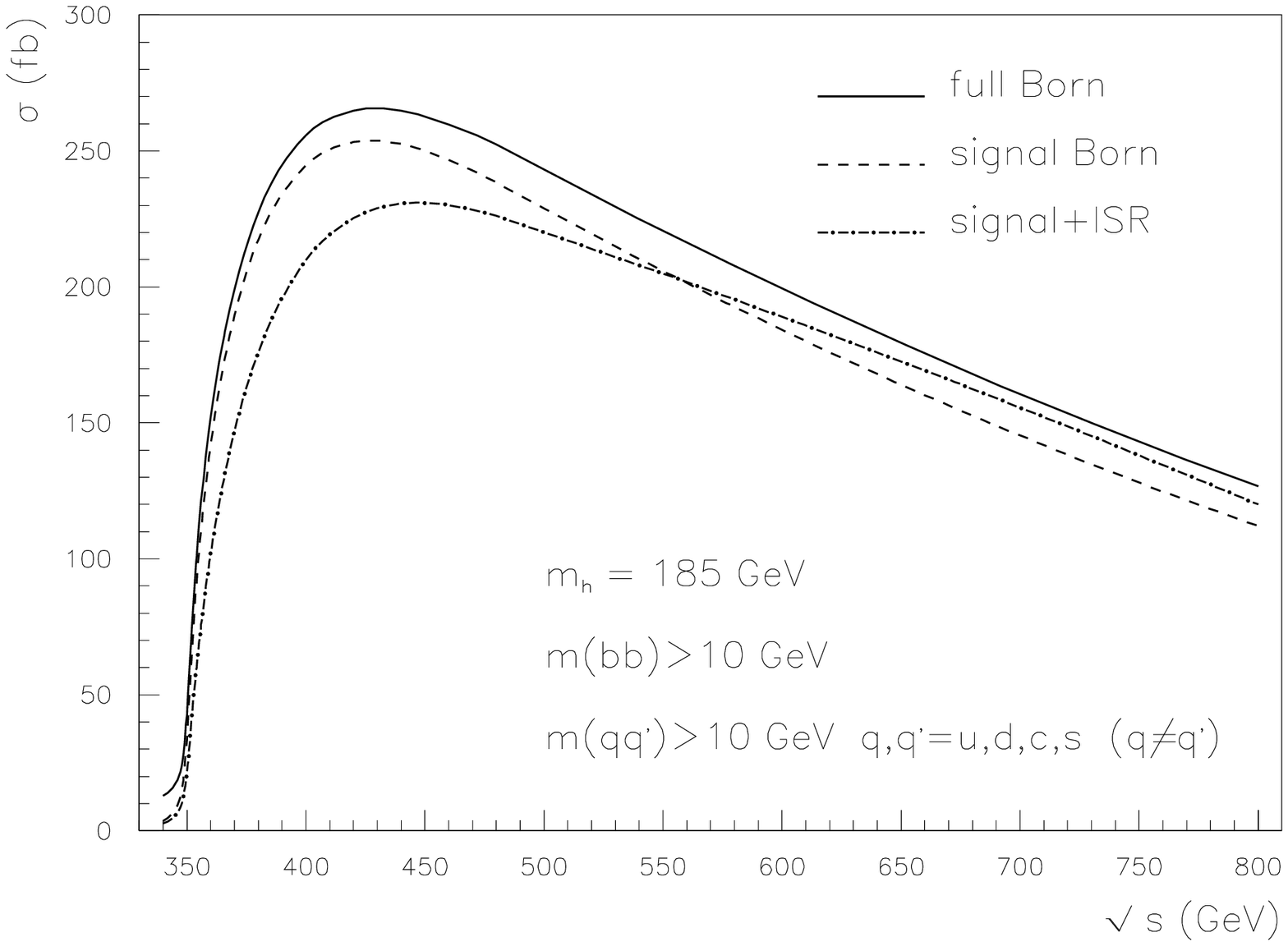,height=8.cm,width=10.cm}
\caption{\small Full six-quark electroweak cross-section (solid line)
and $t\bar t$ signal (dashed line) in the Born approximation, and $t\bar t$
signal with initial-state radiation (dash-dotted line), as a function of the
c.m. energy.}
\label{fig:f185sig}
\ece
\efig
\ece
In Fig.~\ref{fig:signwa} the signal cross-section without kinematical cuts is
plotted together with the cross-section in the narrow-width 
approximation (NWA).
The latter is calculated as the product of the cross-section for
$e^+e^-\to t\bar t$, and of the branching ratios of the decays $W\to q\bar q'$,
assuming the branching ratio of $t\to Wb$ to be exactly unity.
The difference between the two calculations is about $15\%$ in the region
near the threshold, and it decreases, as expected, with increasing c.m. energy:
at 500 GeV it is $3\%$, while at 800 GeV it is less than $1\%$.
These results give a measure of the off-shellness effects connected
with the top-quark and $W$-boson widths.
\bce
\bfig
\bce
\epsfig{file=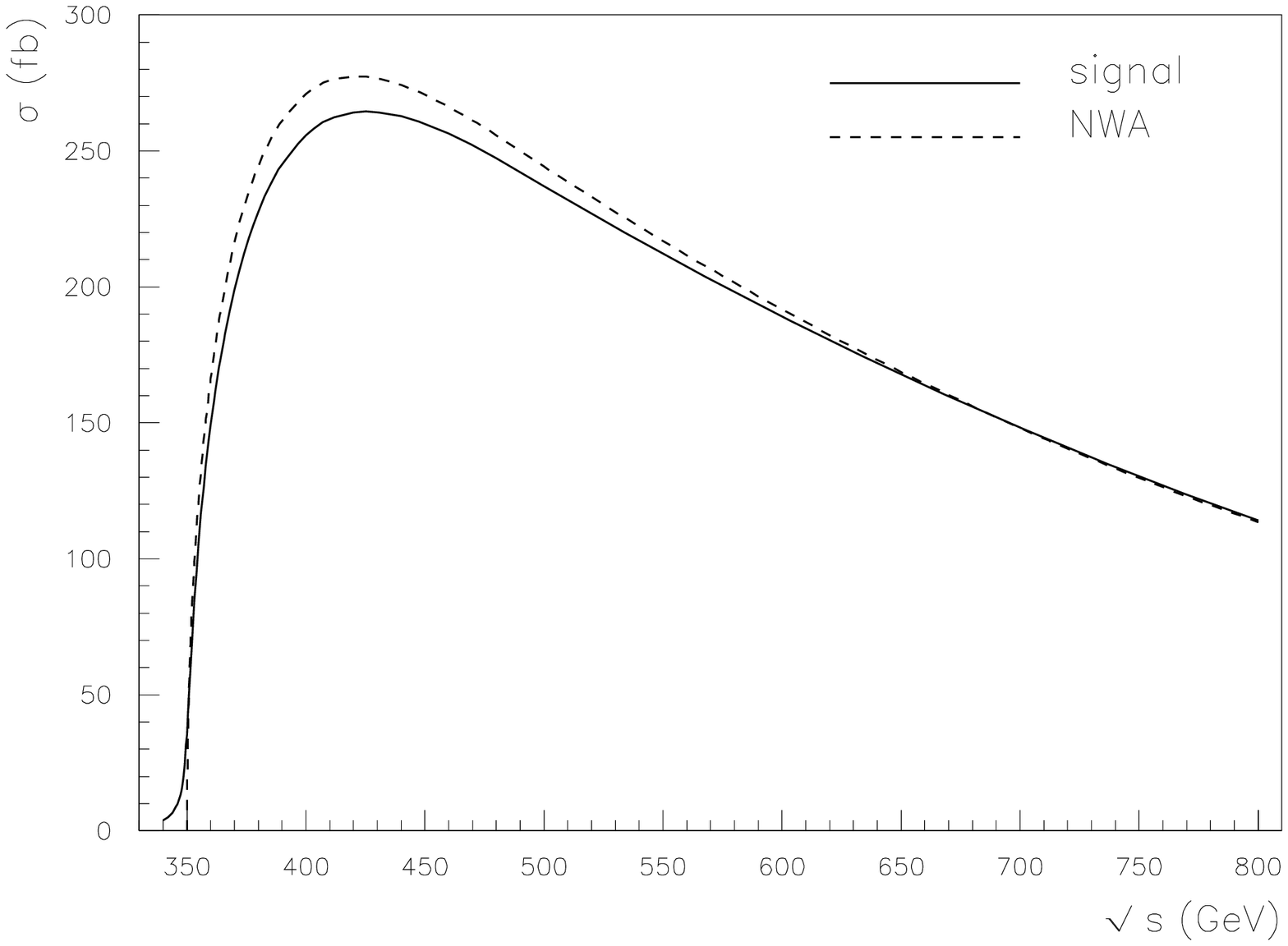,height=8.cm,width=10.cm}
\caption{\small Signal cross-section (solid line) without cuts compared with the
narrow-width approximation (dashed line), as a function of the c.m. energy.}
\label{fig:signwa}
\ece
\efig
\ece
The \xss\ for the two values of the Higgs mass, $m_H=100$ and 185 GeV,
have been found to differ by less than $1\%$ at energies above
the threshold region, while at lower energies, differences of up to $20-30\%$
occur. This is due to the fact that the signal at low energy is not large
enough to hide the Higgs-mass effects. Moreover, such effects decrease with
increasing energy.
In order to make a detailed study of the dependence on the Higgs mass at low
energy, the \xs\ at the threshold for $t\bar t$ production, $\sqrt s=350$ GeV,
has been calculated for various Higgs masses in the range from 100 GeV to 185
GeV. The results are shown in Fig.~\ref{fig:hmscan}, where the \xs\ is plotted
as a function of the Higgs mass. Variations of the order of $10\%$ can be seen
in this plot, which shows 
the importance of complete calculations to keep under
control the background effects and uncertainties that come from 
not knowing 
 the Higgs mass.
\bce
\bfig
\bce
\epsfig{file=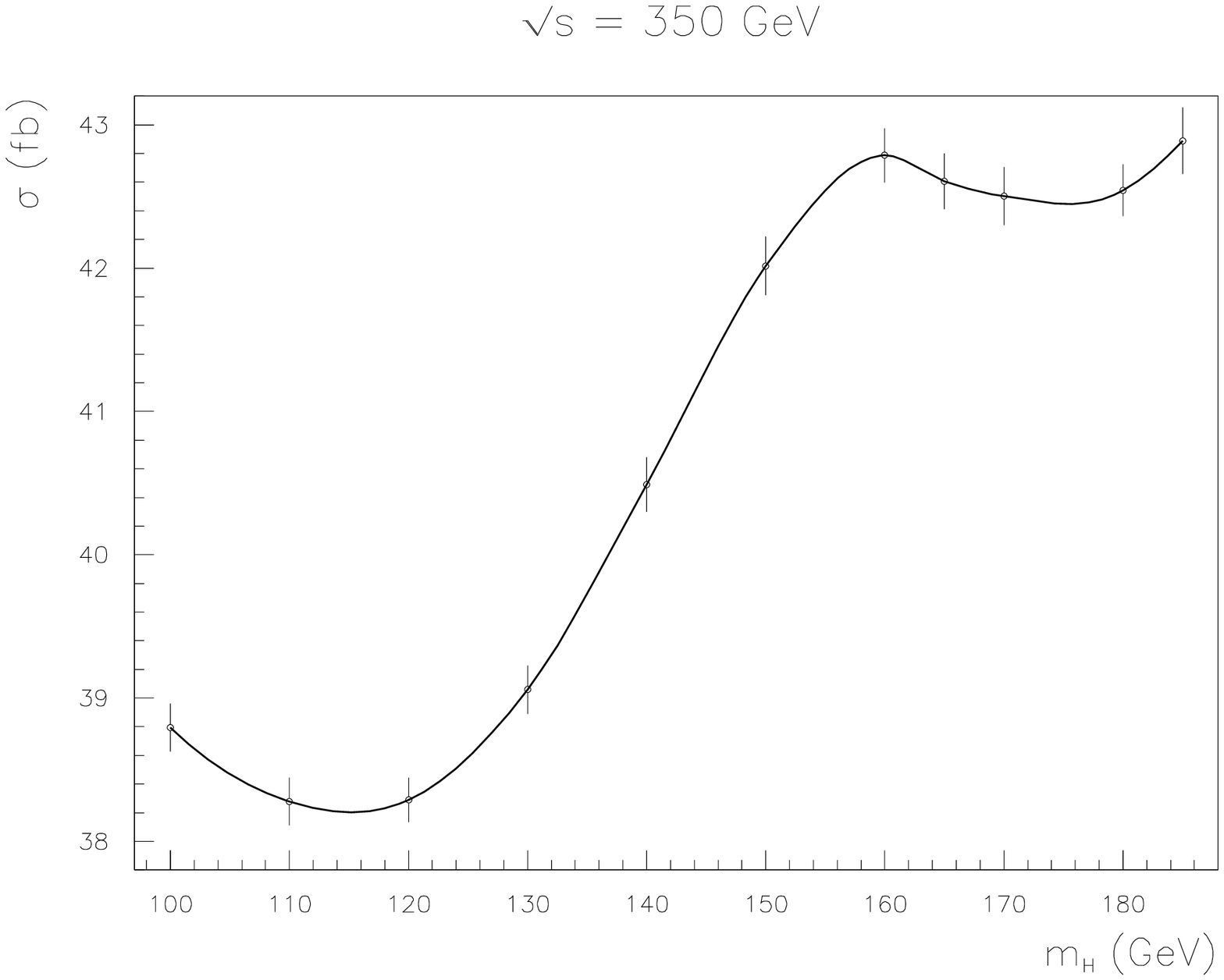,height=9.cm,width=12.cm}
\caption{\small Total cross-section as a function of the Higgs mass at the
threshold for $t\bar t$ production.}
\label{fig:hmscan}
\ece
\efig
\ece

\subsection{Distributions}
\label{subsect:dist}
Two samples of events have been generated at a c.m. energy of 500 GeV
and with a Higgs mass of 185 GeV. One sample is in the Born approximation,
while the other includes {\em ISR} and {\em BS}. The numbers of events, of the order of
$10^5$, have been determined by assuming a luminosity of $500$~fb$^{-1}$,
which is the integrated value expected in one year of run.

In the definition of observable distributions for the class of processes 
considered here, we must take into account the fact that quark flavours 
other than
 $b$ cannot be identified. As a consequence, two kinds of distributions,
labelled  ``exact'' and ``reconstructed'', are considered in the following:
the ``exact'' distributions are calculated by identifying all the quarks; the
``reconstructed'' distributions are calculated by means of the following
algorithm. The momenta $q_1,\ldots ,q_4$ of the four quarks other than $b$
and $\bar b$ are first considered and, for every pair $(q_i,q_j)$, the invariant
mass $m_{ij}=\sqrt{(q_i+q_j)^2}$ is calculated; then the two $W$ particles,
$W_1$ and $W_2$, are reconstructed as the pairs $(q_i,q_j)$ and $(q_k,q_l)$
such that the quantity $|m_{ij} - M_W| + |m_{kl} - M_W|$ is minimized; the 
top-quark is then determined by taking the combination $(b,W_i)$,
$(\bar b,W_j)$, which minimizes the quantity 
$|m_{bW_i} -\tilde m_t| + |m_{\bar b W_j} - \tilde m_t|$, where $\tilde
m_t=175$ GeV is the {\em nominal  top mass}.

The invariant mass of the top-quark is studied in Fig.~\ref{fig:mtop}.
In the plot $(a)$ a comparison is made between the exact (dashed line) and the
reconstructed (solid line) distribution in the Born approximation and a good
agreement can be observed. In order to further check the reconstruction
procedure, in particular the dependence on the adopted value of
$\tilde m_t$,
some tests have been made by taking values in the range $170$ GeV $<\tilde
m_t<180$
GeV and fitting the resulting histograms with Breit--Wigner distributions.
The values of the {\it physical 
top mass} obtained in the various cases are identical, within
the statistical errors.
\bce
\bfig
\bce
\epsfig{file=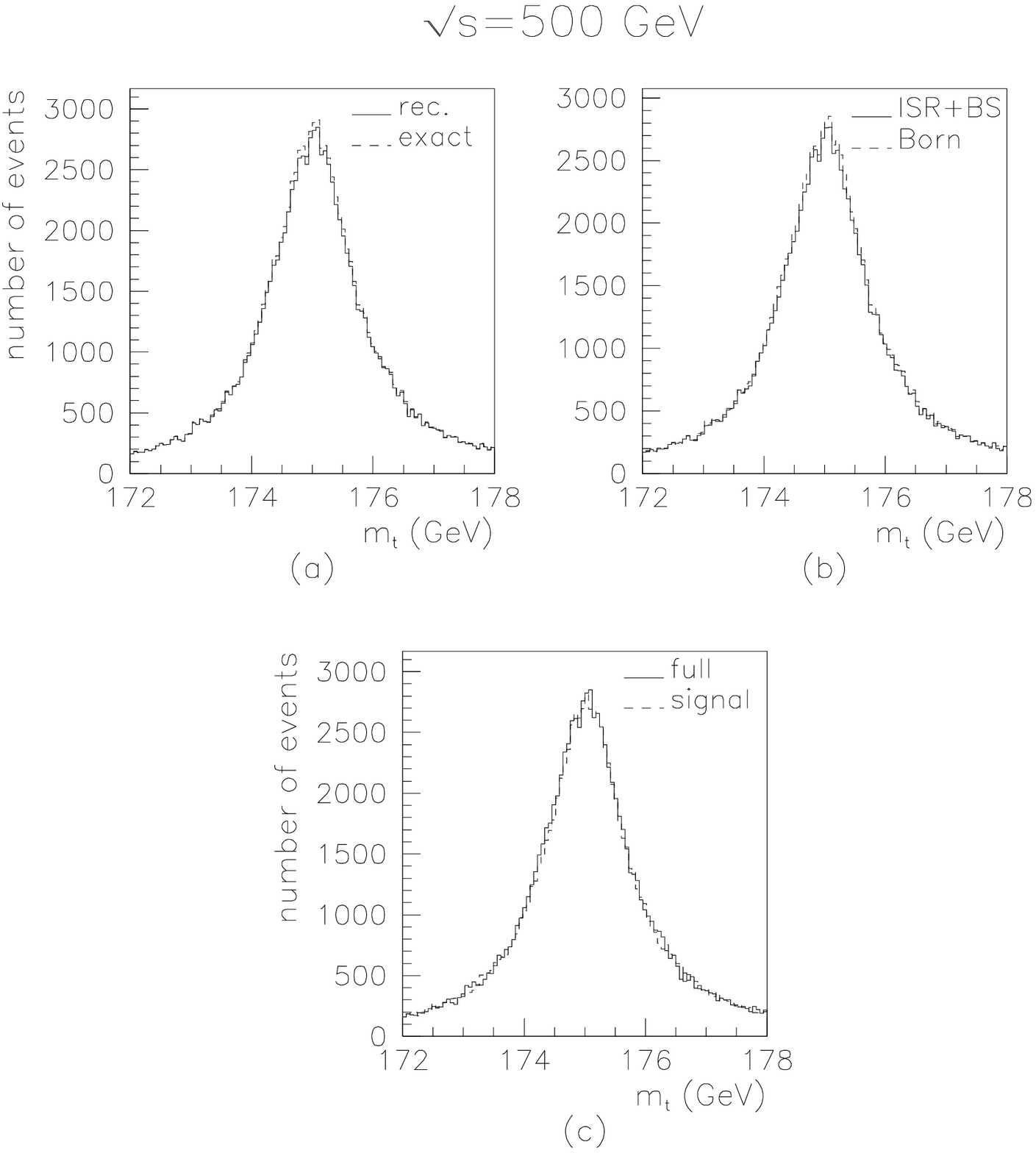,height=14.5cm,width=14.cm}
\caption{\small Invariant mass of the top-quark. $(a)$: exact (solid line)
and reconstructed (dashed line) distribution; $(b)$: distribution in the Born
approximation (dashed line) and with initial-state radiation and beamstrahlung
(solid line); $(c)$: full calculation (solid line) and signal (dashed line).}
\label{fig:mtop}
\ece
\efig
\ece
The radiative effects are shown in the plot $(b)$ of Fig.~\ref{fig:mtop},
for the reconstructed distribution. They do not apparently give a substantial
modification. In the plot $(c)$ the role of background diagrams is
studied, by comparing the result of the full calculation with the signal alone.
The background does not introduce any observable distortion. More quantitative
results have been obtained by means of fits to the histograms with Breit--Wigner
distributions. All the histograms in Fig.~\ref{fig:mtop} give the same value
of $m_t$, so that it can be safely concluded
 that electroweak backgrounds, as well
as {\em ISR} and {\em BS} do not give any bias in the determination of the physical mass
via the direct reconstruction method on the scale of precision of 100 MeV.

The angular distribution of the top-quark with respect to the beam axis
is a good indicator of the spin nature and of the couplings of the top-quark.
 This variable is illustrated in Fig.~\ref{fig:ctop}. As in the case of
the invariant mass, the exact and reconstructed distributions  have been
checked to be in very good agreement.
It should be observed that radiative and background effects are of the same
magnitude (in particular the former are dominated by the {\em ISR}). The shapes of
the histograms are in good qualitative agreement with the angular distribution
predicted by the lowest-order analytic calculation for
the process of real $t\bar t$ production.
\bce
\bfig
\bce
\epsfig{file=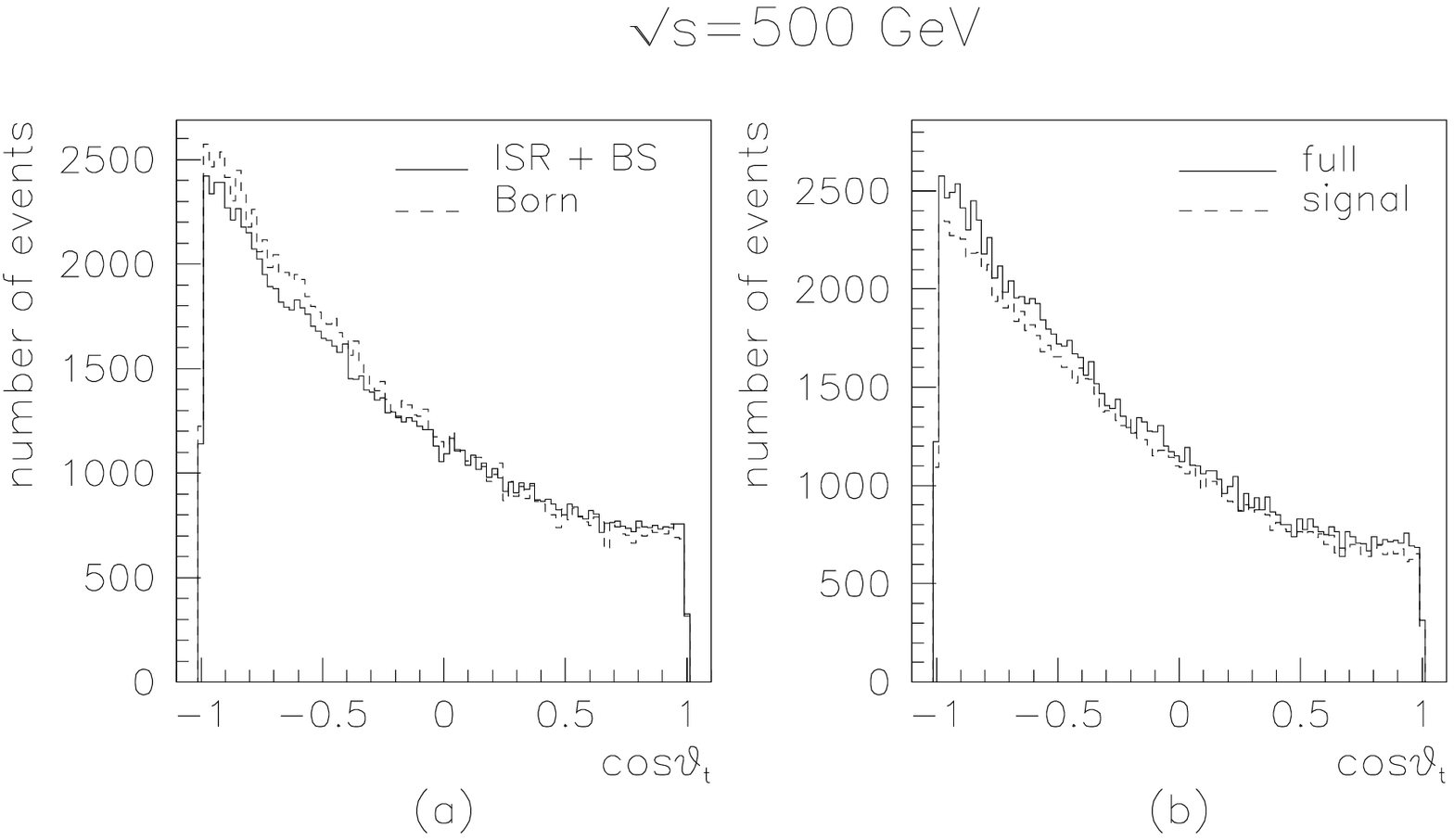,height=7.cm,width=13.cm}
\caption{\small Reconstructed angular distribution of the top-quark with
respect to the $e^+$ beam axis. $(a)$: results in the Born approximation (dashed
histogram) and with initial-state radiation and beamstrahlung (solid histogram).
$(b)$: full calculation (solid line) and signal contribution (dashed line).}
\label{fig:ctop}
\ece
\efig
\ece

The most effective way to obtain a separation between the $t\bar t$ signal
and the QCD backgrounds, as already pointed out by some
authors~\cite{sixj,desy123a}, is to analyse event-shape variables, such as
thrust~\cite{thrust}, sphericity~\cite{spheri}, spherocity~\cite{sphero},
$C$ and $D$ parameters~\cite{cdpar}, etc.
A comparison between pure QCD ($O(\alpha_{em}^2\alpha_s^4)$) six-jet events and
the $t\bar t$ signal has been performed in ref.~\cite{sixj} for the thrust and
sphericity distributions, and other shape variables have been studied in the
same article for the QCD contributions only.
In the present work several such variables have been analysed for the
electroweak contributions, and the effects of the electroweak backgrounds
and of {\em ISR} and {\em BS} have been studied.
The thrust and $C$ parameter distributions for the process under consideration
are shown in Fig.~\ref{fig:shape}. In the upper row the radiative effects are
displayed, while in the plots of the lower row the signal is compared with the
full result. In the radiative case, the distributions are calculated after
going to the c.m. frame.
Remarkable effects due to {\em ISR} and {\em BS} can be seen in these plots and in
particular in the thrust distribution, where the peak is strongly reduced with
respect to the Born approximation and the events are shifted towards the lower
values of $T$, which correspond to spherical events. It is interesting to
observe that this phenomenon is of help for the selection of the signal with
respect to QCD backgrounds.
From the plots in the second row, it can be seen that the presence of the
electroweak backgrounds, although visible, is almost negligible for both
observables.
\bce
\bfig
\bce
\epsfig{file=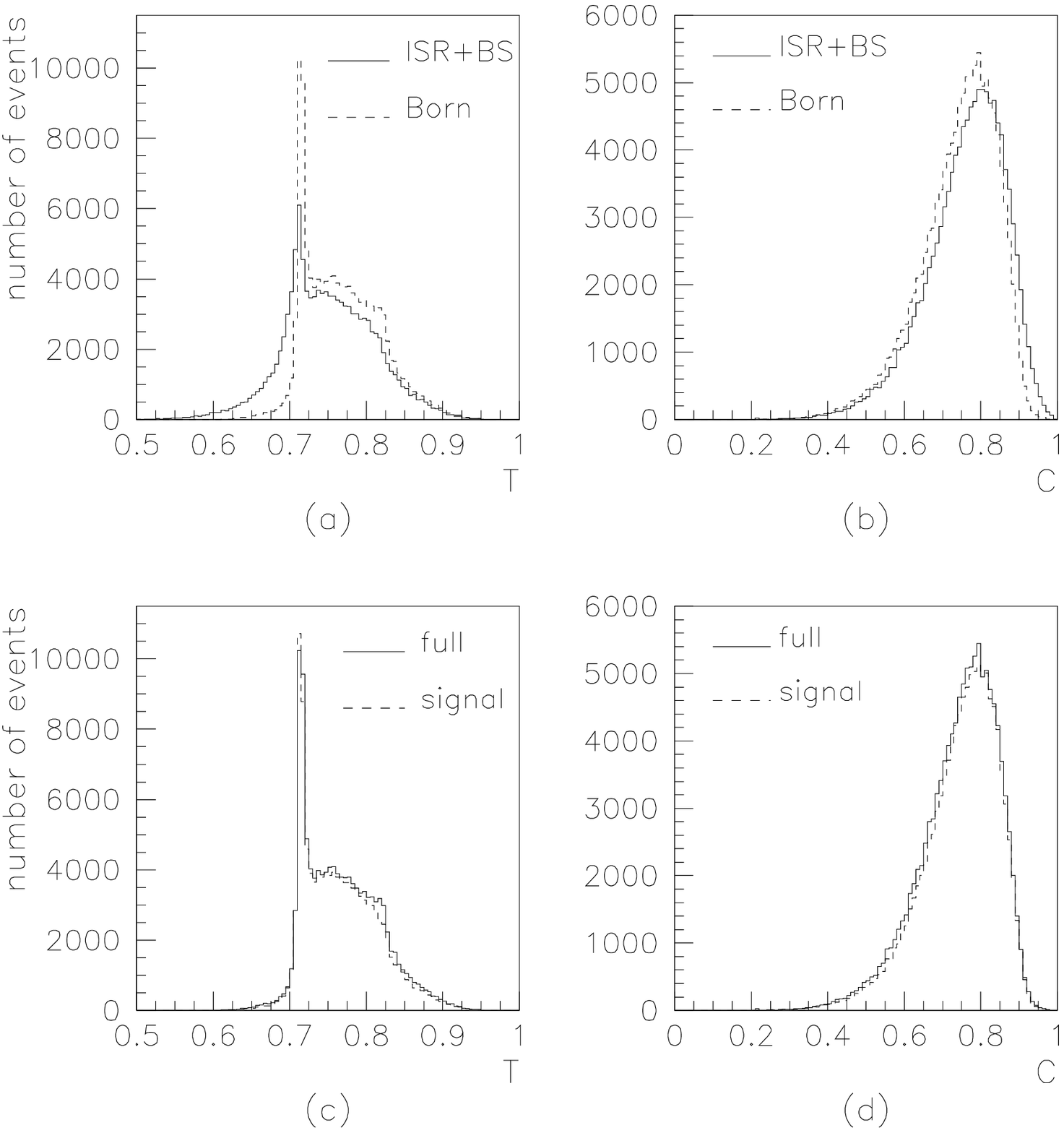,height=14.cm,width=14.cm}
\caption{\small Event-shape variables. $(a)$: thrust distribution in the Born
approximation (dashed histogram) and with initial-state radiation and
beamstrahlung (solid histogram); $(b)$: $C$ parameter distribution, as in $(a)$;
$(c)$: thrust distribution in the Born approximation from the full calculation
(solid histogram) and from the signal contributions alone (dashed histogram);
$(d)$: $C$ parameter distribution, as in $(c)$.}
\label{fig:shape}
\ece
\efig
\ece
The remarkable change of the thrust distribution after inclusion of radiation
can be better understood by observing the dependence of this distribution on
the c.m. energy, which is analysed in Fig~\ref{fig:thrust}, where four samples
of 10000 events each, at the energies of 360, 500, 800 and 1500 GeV, are
studied. The peak structure that is present at 500 GeV is completely lost at
360 GeV, and this explains the lowering of the peak at 500 GeV in the presence
of {\em ISR} and {\em BS}, as this reduces the available c.m. energy.
At 800 and 1500 GeV the peak is shifted towards the collinear region $T\sim 1$,
as a consequence of the Lorentz boost of the $t$ and $\bar t$ quarks.
\bce
\bfig
\bce
\epsfig{file=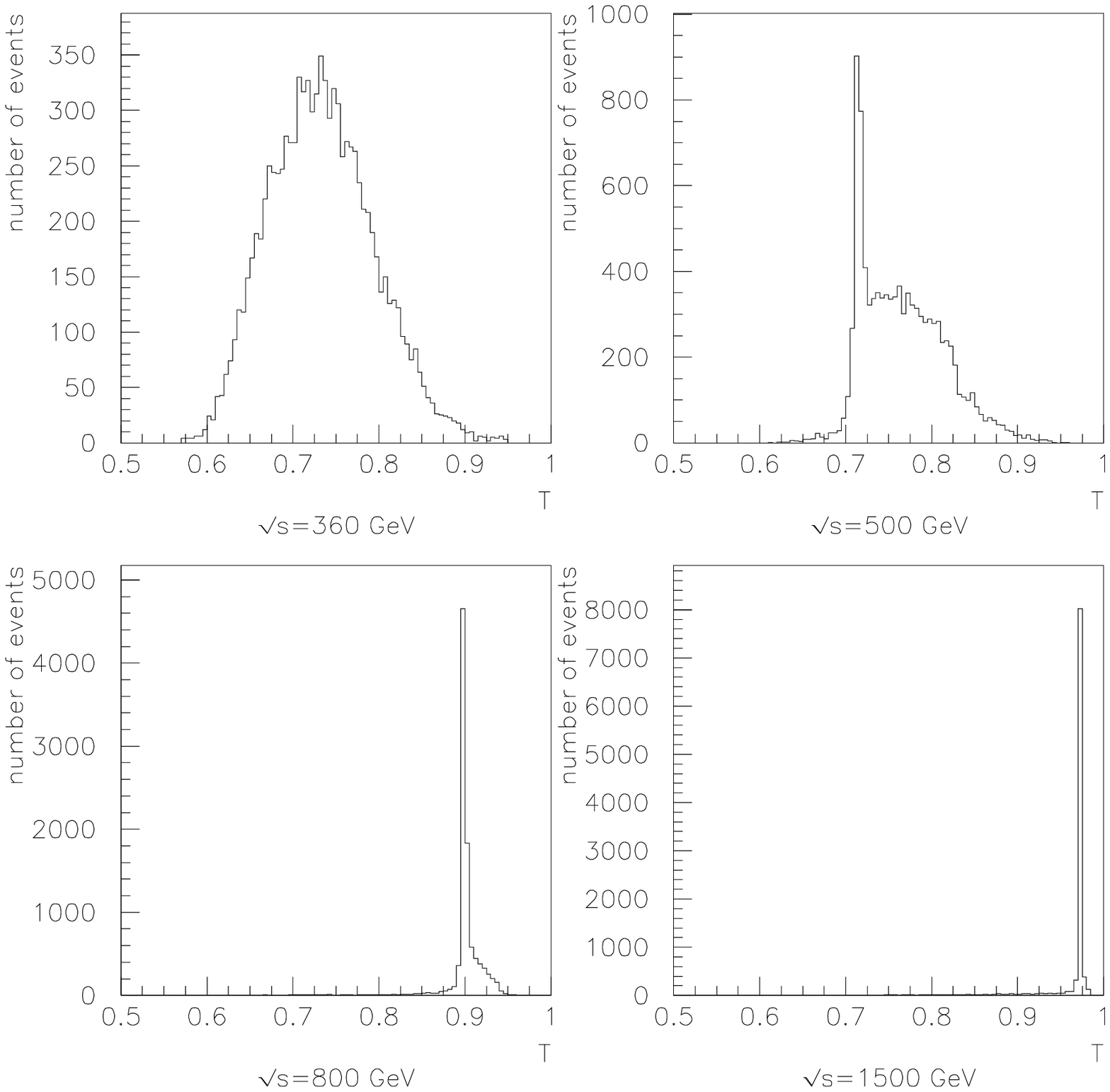,height=14.cm,width=14.cm}
\caption{\small Thrust distribution in the Born approximation at 360, 500, 800
and 1500 GeV.}
\label{fig:thrust}
\ece
\efig
\ece
As a conclusion, we can say that, at 500 GeV, in view of the results of the
pure QCD processes, studied in ref.~\cite{sixj}, the thrust variable is the
most effective in discriminating pure QCD backgrounds, also in the presence of
electroweak backgrounds and of {\em ISR} and {\em BS}. At higher energies this separation
appears to be more and more problematic.

On the other hand, the backgrounds of $O(\alpha_{em}^4\alpha_s^2)$, given by
$2\to 4$ processes with subsequent gluon emission from a quark line, should be
considered (a study of contributions of this class for semi leptonic signatures
is made in ref.~\cite{4jln}, but without an analysis of event-shape variables).
A rough estimate of the leading contributions of this kind could be
obtained by considering a four-fermion process of the form $e^+e^-\to W^+W^-\to
4$ jets, similar to what is done in ref.~\cite{desy123a}. A test made by means
of the four-fermion program WWGENPV has confirmed the results of
ref.~\cite{desy123a} for the thrust and has led to similar conclusions for the
$C$ parameter: such processes appear to be well separated from the top-quark 
signal and thus appear to be less dangerous than the pure QCD backgrounds.

\section{Conclusions}
\label{sect:concl}

The production of $t\bar t$ pairs has been studied in processes with six quarks
in the final state, at the energies of the NLC. The signatures
considered contain one $b\bar b$ pair, and receive contributions from both
charged and neutral currents. The top-quark signal is present only in the
charged-current terms. The purely electroweak contributions have been
considered and complete tree-level calculations have been performed.

The \xs\ has been calculated by means of a computer program already used for
other phenomenological studies on $6f$ processes and adapted here to
sample the new diagram topologies.
The importance of the electroweak  backgrounds and of the off-shellness effects
has been examined. Above the threshold for $t\bar t$ production, the former
are of the order of several per cent and the latter are at the per cent
level. Near the threshold, both effects are sizeable and, in particular, a
study of the dependence of the \xs\ on the Higgs mass at threshold
shows that variations of the order of 10$\%$ occur for Higgs masses between
100 GeV and 185 GeV. A complete calculation is needed to keep such effects
under control
and to have a $1\%$ accuracy.

Some distributions have been studied in a realistic approach, by using a
reconstruction algorithm for the top-quark that takes into account the
impossibility of identifying quark flavours other than $b$. The
invariant mass of the top-quark has thus been studied and the presence
of electroweak background contributions as well as the initial-state radiative
effects have been found not to affect the determination of the mass on the
scale of experimental precision expected at NLC.

The angular distribution of the top-quark with respect to the beam axis,
which is directly related, in the case of real production, to the quantum
numbers of the top-quark, has been calculated and shown to be in
qualitative agreement with the expectation suggested by the real production
case.

Finally, some event-shape variables have been studied. At a c.m. energy of 500
GeV the thrust distribution turns out to be the most interesting for the aim of
discriminating the leading QCD backgrounds, as suggested by other authors who
discussed the top-quark signal alone. The effects of electroweak
backgrounds and of {\em ISR} and {\em BS} have been shown here not to alter these
conclusions. At higher energies, the Lorentz boost gives to the event a more
collinear shape, so that the separation of QCD backgrounds could become more
difficult.

The study presented in this paper has been performed by means of a computing
program that can equally well deal with semi leptonic signatures and can be
switched in a straightforward manner to treat polarized scattering. Moreover,
by employing the new version of ALPHA~\cite{alpha1}, which embodies also the
QCD Lagrangian, complete strong and electroweak results could be obtained.

\vspace{1.truecm}
\noindent
{\bf Acknowledgements}\\
F.~Gangemi thanks the INFN, Sezione di Pavia,
for the use of computing facilities. The work of M.~Moretti is funded
by a Marie Curie fellowship (TMR-ERBFMBICT 971934).


\begin{thebibliography}{99} 
%

\bibitem{lc}{E.~Accomando et al., \prp\ {\bf 299} (1998) 1;\\ F.~Boudjema, 
             Pramana {\bf 51} (1998) 249.}

\bibitem{lhc}{G.~Jarlskog and D.~Rein (eds.), Proceedings of the ``Large Hadron
             Collider Workshop'', Aachen, 1990;\\
             ATLAS Technical Proposal, CERN/LHC/94-43 LHCC/P2 (December 1994);
             CMS Technical Proposal, CERN/LHC/94-43 LHCC/P1 (December 1994).}

\bibitem{Fermilab}{P.~Bhat, H.~Prosper and S.~Snyder, Int. J. Mod. Phys. 
                   {\bf A13} (1998) 5113.}

\bibitem{kek}{F.~Yuasa, Y.~Kurihara and S.~Kawabata,
              \plb{\bf 414} (1997) 178.}

\bibitem{to1}{E.~Accomando, A.~Ballestrero and M.~Pizzio, \npb{\bf 512} (1998)
             19; in R.~Settles (ed.) ``$e^+e^-$ Linear Colliders: Physics and
             Detector Studies, Part E'', DESY 97-123E, p.~31;
             {\tt hep-ph/9807515};\\
             A.~Ballestrero, Acta Phys. Pol. {\bf B29} (1998) 2811.}

\bibitem{sixfzpc}{G.~Montagna, M.~Moretti, O.~Nicrosini and F.~Piccinini,
                  Eur. Phys. J. {\bf C2} (1998) 483.}

\bibitem{gmmnp97}{F.~Gangemi, G.~Montagna, M.~Moretti, O.~Nicrosini and
                 F.~Piccinini in R.~Settles (ed.) ``$e^+e^-$ Linear Colliders: 
                 Physics and Detector Studies, Part E'', DESY 97-123E, p. 393.}

\bibitem{6fhiggs}{F.~Gangemi, G.~Montagna, M.~Moretti, O.~Nicrosini and
                 F.~Piccinini, {\tt hep-ph/9811437}, to appear in
                 Eur. Phys. J. {\bf C}.}

\bibitem{keystone}{T. Ohl, talk given at ``Workshop on Physics and Detectors
                   for Future $e^+e^-$ Linear Colliders'', 26-29 September 1998,
                   Keystone (CO).}

\bibitem{4jln}{S.~Moretti, {\tt hep-ph/9901438}.}

\bibitem{threshold}{V.S.~Fadin and V.A.~Khoze, Sov. J. Nucl. Phys. {\bf 48}
 (1988) 309;\\ M.J.~Strasseler and M.~Peskin, \prd {\bf 43} (1991) 1500.}

\bibitem{teub}{T.~Teubner, {\tt hep-ph/9904243},
 talk given at the ``Cracow Epiphany Conference on Electron-Positron
 Colliders'', Cracow, 1999, and refs. therein.}

\bibitem{sixj}{S.~Moretti, \plb{\bf 420} (1998) 367; \npb {\bf 544} (1999) 289.}

\bibitem{alpha}{F.~Caravaglios and M.~Moretti, Phys.~Lett. {\bf B358}
               (1995) 332.}

\bibitem{higgspv}{Program {\tt HIGGSPV},  
      by G.~Montagna, O.~Nicrosini and F.~Piccinini;  write up
      in~\cite{dpeg,wweg}.} 
 
\bibitem{wwgenpv}{Program  {\tt WWGENPV},  
      by G.~Montagna, O.~Nicrosini and  F.~Piccinini; write  up in 
      \cite{dpeg,wweg}. See also\\  
      G.~Montagna, O.~Nicrosini  and  F.~Piccinini, Comput. Phys. Commun. {\bf 90} (1995) 141;\\ 
      D.G.~Charlton, G.~Montagna, O.~Nicrosini and F.~Piccinini,  Comput. Phys. Commun.  
      {\bf 99} (1997) 355. } 
 
\bibitem{sf}{E.A.~Kuraev and V.S.~Fadin, Sov. J. Nucl. Phys. {\bf 41} (1985) 
             466;\\ G.~Altarelli and G.~Martinelli, in {\it Physics at LEP},
             J.~Ellis and R.~Peccei, eds., CERN Report 86-02
             (Geneva, 1986), vol.~1, p.~47;\\
             O.~Nicrosini and L.~Trentadue, Phys. Lett. {\bf B196} (1987) 551,
             Z. Phys. {\bf C39} (1988) 479;\\
             F.A.~Berends, G.~Burgers and W.L.~van Neerven, \npb{\bf 297} (1988)
             429.}
 
\bibitem{circe}{T.~Ohl, Comput. Phys. Commun. {\bf 101} (1997) 269.} 
 
\bibitem{hwg}{M.~Carena, P.M.~Zerwas (convenors), ``Higgs Physics",
                in~\cite{lep2}, vol.~1, p.~351.} 
 
\bibitem{kniel}{B.~Kniehl, Phys. Lett. {\bf B244} (1990) 537. } 

\bibitem{desy123a}{P.~Igo-Kimenes (convener) and J.H.~K\"uhn (advisor) in
  Proceedings of the Workshop on ``$e^+e^-$ Collisions at 500 GeV. The Physics
  Potential'', Munich, Annecy, Hamburg, ed. P.M.~Zerwas, DESY 92-123, 1992,
  part A, p.~327.}

\bibitem{thrust}{S.~Brandt, Ch.~Peyrou, R.~Sosnowski and A.~Wroblewski,
                 Phys. Lett. {\bf 12} (1964) 57;\\
                 E.~Farhi, \prl {\bf 39} (1977) 1587.}

\bibitem{spheri}{J.D.~Bjorken and S.~Brodsky, \prd {\bf 1} (1970) 1416.}

\bibitem{sphero}{H.~Georgi and M.~Machacek, \prl\ {\bf 39} (1977) 1237.}

\bibitem{cdpar}{G.~Parisi, \plb {\bf 74} (1978) 65;\\
          J.F.~Donoghue, F.E.~Low and S.Y.~Pi, \prd {\bf 20} (1979) 2759;\\
          R.K.~Ellis, D.A.~Ross and A.E.~Terrano, \npb {\bf 178} (1981) 421.}

\bibitem{alpha1}{F.~Caravaglios, M.L.~Mangano, M.~Moretti and R.~Pittau,
                 \npb {\bf 539} (1999) 215.}
 
\bibitem{lep2}{  G.~Altarelli, T.~Sj\"ostrand and 
      F.~Zwirner, eds., {\it Physics at LEP2}, CERN Report  96-01
 (Geneva, 1996), vols.~1 and  2.} 

\bibitem{dpeg}{M.L.~Mangano, G.~Ridolfi (convenors), ``Event Generators for  
                Discovery Physics'', in~\cite{lep2}, vol.~2, p.~299.} 

\bibitem{wweg}{D.~Bardin, R.~Kleiss (convenors), ``Event Generators for $WW$  
              Physics'', in~\cite{lep2}, vol.~2, p.~3.} 

\end{thebibliography}
\end{document}